# A study of stability analysis of pyroclastic covers based on electrical resistivity measurements

## Resistivity measurements for stability analysis

R. Di Maio, E. Piegari

*Department of Earth Sciences, University of Naples "Federico II", Largo S. Marcellino 10, 80138 Napoli, Italy*

E-mail: esterpiegari@gmail.com

**Abstract**

Usually, the degree of stability of a slope is quantified by the Factor of Safety whose values depend on physical and mechanical soil properties analyzed on samples of much reduced sizes or referring to very small soil volumes around porous probes. To overcome the limit of point-sampled information, we propose a semi-empirical approach based on the use of geophysical methods and the employment of a geophysical Factor of Safety recently introduced by the authors in terms of local resistivities and slope angles. In this paper, we show an application of our proposal on a test area of about 2000 m$^2$ on Sarno Mountains (Campania Region – Southern Italy), where shallow landslides involving pyroclastic soils periodically occur triggered by critical rainfall events. Starting from two resistivity tomography surveys performed on the test area in autumn and spring, we obtained maps of the geophysical Factor of Safety at different depths for the two seasons. We also estimated the values of the Factor of Safety by using the infinite slope model in the dry and saturated scenario. A comparison between the values of the geophysical and geotechnical Factor of Safety shows advantages and disadvantages of our approach.

## 1. Introduction

During the last twenty years, mainly two kinds of models have been developed to forecast the occurrence of rainfall-induced shallow landslides: *empirically based* and *physically based models*.

---

\* Corresponding author.
 *E-mail address*: esterpiegari@gmail.com (E. Piegari).

The *empirically based models* typically identify rainfall intensities and duration required to trigger landslides by a cause-effect relationship between landslide occurrence and rainfall recorded at rain gauge stations (Dai et al., 2002; Hong and Adler, 2007). In this respect, a recent review (Guzzetti et al., 2007) has pointed out the large variety of variables used to define empirical rainfall thresholds (more than 20), which demonstrate the lack of shared criteria. In addition, different thresholds have been proposed for similar and even for the same geographical areas. For instance, for shallow landslides involving pyroclastic soils of the investigated area (see below) the regional threshold proposed by Guadagno (1991) is higher and steeper than the threshold proposed by Calcaterra et al. (2000) for all landslide types of the same region.

The *physically based models* estimate the amount of precipitation needed to trigger slope failures by using slope stability analyses combined with infiltration models (Sidle, 1992; Montgomery and Dietrich, 1994; Dietrich et al., 1995, Wu and Sidle, 1995; Iverson, 2000; Baum et al., 2002; Lu and Godt, 2008; Godt et al., 2009). In this case, the degree of stability is quantified by a dimensionless parameter known as the Factor of Safety, whose values depend on soil properties, like friction angle, suction, unit weight, which are obtained by analyzing samples of much reduced sizes or referring to very small soil volumes around porous probes.

Both these approaches show limitations in required spatial information on the hydrological, lithological, morphological and soil characteristics that control landslide triggering. To overcome the limit of point-sampled information, i.e. very small rain water collecting area (of the order of $cm^2$, for a rain gauge) and very small soil volumes around porous probes, we propose the use of geophysical methods in slope stability analysis. Actually, in the last decades the use of geophysical methods for landslide investigations is greatly increased (Hack, 2000; Jongmans and Garambois, 2007; Chambers et al., 2011). However, these methods do not provide quantitative information on the stability analysis since geophysical parameters are not directly linked to the geological and mechanical properties required by geologists and engineers. Recently, we have proposed the first attempt to relate the Factor of Safety to in situ-(geo)physical quantities (Piegari et al., 2009a). The proposed geophysical Factor of Safety, which is a function of electrical resistivity and slope angle, should not be considered as exhaustive, since it is just the first step of approximation of an analytical function that we postulate would depend on many in-situ measurable physical quantities, which will be analyzed in future work.

In this paper, we use the proposed geophysical Factor of Safety to analyze the stability of a test area of about 2000 $m^2$ located at Mt. Pizzo d'Alvano (Salerno – Southern Italy) above the initial detachment areas of two landslides that occurred on the $5^{th}$ and $6^{th}$ of May 1998 (figure 1). In particular, from two resistivity tomography surveys performed on the test area during the autumnal and spring seasons (Di Maio and Piegari, 2011), we obtain maps of the geophysical Factor of Safety at different depths for the two seasons. We discuss limits and advantages of the proposed approach by comparing the obtained distributions of the geophysical Factor of Safety with those retrieved by the 1D Factor of Safety commonly used in the slope stability analysis of shallow landslides.

* Corresponding author.
  *E-mail address*: esterpiegari@gmail.com (E. Piegari).

## 2. Geoelectrical characterisation of the pyroclastic cover on Mt. Pizzo d'Alvano (Salerno, Italy)

The pyroclastic soils under analysis originate from the different explosive phases of the Mt. Somma-Vesuvius volcano. They cover the limestone relief of Mt. Pizzo d'Alvano (1133 m a.s.l.), which is located at about 10 km to east of the volcano and belongs to the Sarno Mountain complex. Since the calamitous event occurred on May 5-6, 1998, which caused 161 casualties and serious damage in the piedmont towns of Sarno, Quindici, Bracigliano and Siano, many laboratory analyses have been performed to characterize the index properties of such ash-fall pyroclastic deposits (Esposito and Guadagno, 1998; Cascini et al., 2000; Crosta and Dal Negro, 2003; Fiorillo and Wilson, 2004; Guadagno and Revellino, 2005; De Vita et al., 2006; Cascini et al., 2008; De Vita et al., 2011).

The thickness of the pyroclastic cover mainly ranges from 2 to 5 meters and lies on a highly fractured carbonate bedrock with open joints filled by pyroclastic soil. The basic stratigraphical sequence of the cover can be roughly characterized by three pyroclastic layers: an upper ashy layer, characterized by the largest grain sizes (*B* horizon), an intermediate pumice level (*C* horizon) with the largest values of porosity, and a lower clayey ash layer that presents the smallest grain sizes (*Bb* - *Bb*$_{basal}$ horizons) (de Riso et al., 1999; Calcaterra et al., 2003; Cascini et al., 2003; De Vita et al., 2006; De Vita et al., 2011).

Recently, we have provided an electrical characterization of these pyroclastic deposits (Di Maio and Piegari, 2011) in the attempt to estimate their highly variable water content by carrying out resistivity tomography (ERT) surveys in a test area of about 2000 m$^2$ located above the initial detachment areas of two landslides that occurred on the 5$^{th}$ and 6$^{th}$ of May 1998 at Mt. Pizzo d'Alvano (figure 1). Specifically, a 2D ERT survey along nine parallel profiles 58 m long and distant each other 4 m (see figure 1) was performed during both the autumnal and spring seasons. The data were collected using an IRIS-SYSCAL PRO System, with multielectrode cable characterised by automatic switching electrodes spaced of 2 m along the cable. As we needed both reasonably good horizontal and vertical resolution, a Wenner-Schlumberger array with overlapping data levels was used (Loke, 2002), with the minimum and maximum current electrode separation of 6 m and 44 m, respectively, and the minimum potential electrode separation of 2 m. For each line, a total number of 167 measurements and an investigation depth of about 10 m below ground level (b.g.l.) were attained. The collected apparent resistivity data were inverted by the RES3DINV algorithm (Loke, 2002; Loke and Barker, 1996; Loke and Dahlin, 2002) using the finite element method to link the model parameters to the 3D model response and the complete Gauss-Newton technique to determine the change in the model parameters. The rough surface topography characterizing the profiles, ranging from about 738 to 748 m a.s.l., was incorporated into the model. A RMS misfit error of 10.70 and 10.78 resulted for the inverted models of the resistivity data acquired, respectively, in the autumnal and spring seasons, with a percentage of outliers discarded of about 1.7% for both surveys. In figure 2, the volumetric resistivity distribution obtained by the 3D inversion of the 2D ERT data collected in the autumnal season is reported from different points of view.

* Corresponding author.
  *E-mail address*: esterpiegari@gmail.com (E. Piegari).

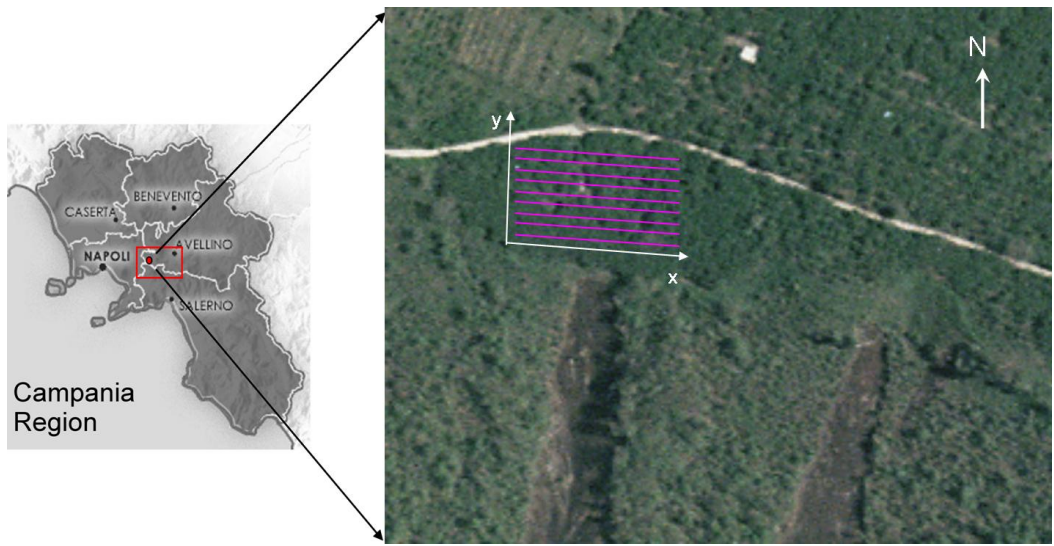

**Figure 1** Map of the survey area. Location of the geophysical investigation (magenta lines) on an orthophotograph of the northern slope of Pizzo d'Alvano. The initial detachment areas of two landslides occurred on the 5$^{th}$ and 6$^{th}$ of May 1998 are also visible.

From the 3D images with horizontal and transversal cuts out (figures 2a and 2b), emerges a clear trend in the dependence of the resistivity with the depth throughout most of the section. After resistivity reaches the minimum value ranging from 100 to 500 Ωm at about 4-5 m b.g.l, it starts to increase going downwards. The increase of the resistivity values on the deep part of the sections clearly shows the interface between the volcanoclastic deposits and the underlying highly fractured carbonate bedrock. Interestingly, from the bottom view (figure 2c) a sharp discontinuity of the resistivity values appears approximately in the middle of the area, which we interpret as a step in the underlying bedrock structure (sketched with black lines in figure 2). Then, also on the basis of geological and geotechnical classifications of the investigated soils (Schmidt, 1981; De Vita et al., 2006), we have characterized the pyroclastic cover through the following different (geo)electrical levels (Di Maio and Piegari, 2011):

- the shallow resistive layer, with about two-meter-thick and characterized by resistivity values ranging from about 1000 to 3500 Ωm, characterizes a pyroclastic deposit subject to a pedogenesis process in its upper part (*B* horizon), and including weathered pomiceous lapilli in its lower sector (*C* horizon);
- the most conductive block, located in a depth range from 2 to 5 m b.g.l. and characterised by resistivity values varying in the range 150÷1200 Ωm, which could be associated with strongly weathered ash materials, represents the overlapping of different eruptive deposits along with pedogenic processes built up (*Bb* and *Bb$_{basal}$* horizons);

\* Corresponding author.
  *E-mail address*: esterpiegari@gmail.com (E. Piegari).

- the resistive deep pattern, observed from a depth of about 5 m b.g.l. and described by an increase of the resistivity values, corresponds to a fractured carbonate basement whose upper part is fractured and filled by finer pyroclastic material coming from the overlying deposit.

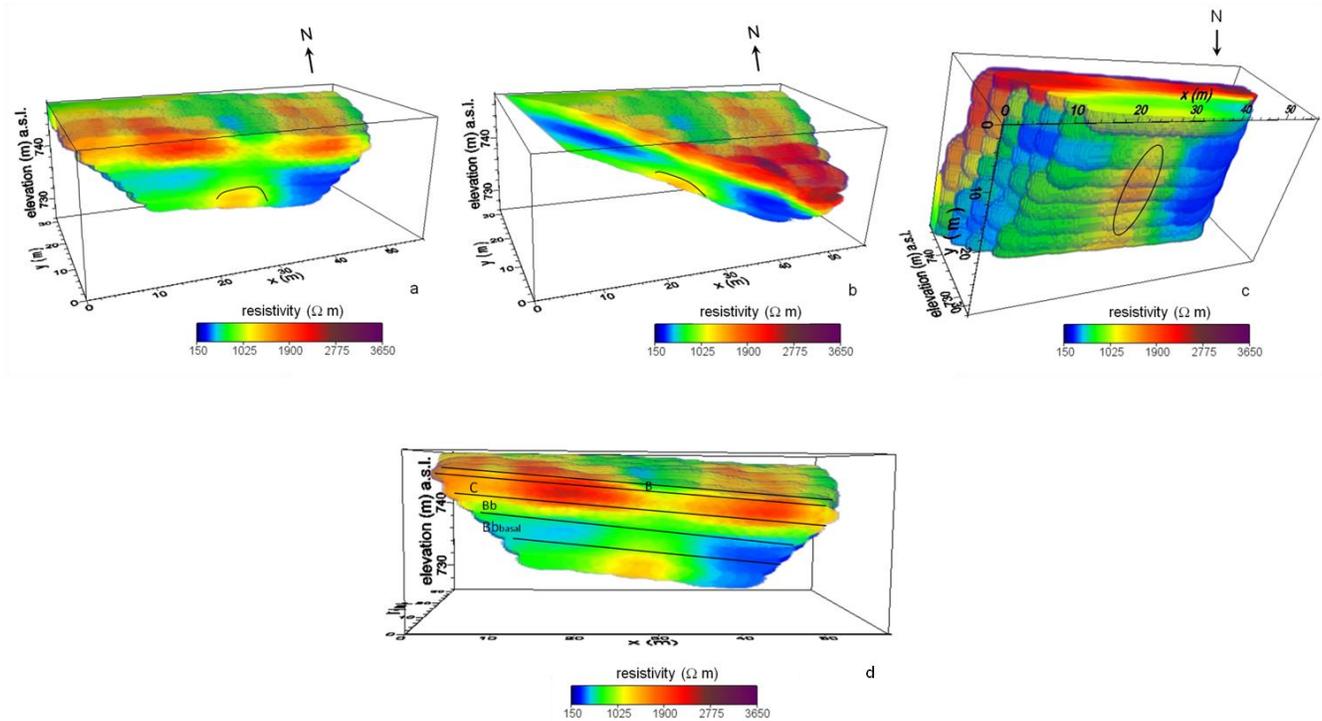

**Figure 2** Resistivities of the investigated buried volume obtained from the results of the 3D data inversion. The volume data are reported with a clipping plane parallel to the direction of the profiles (a) and transversal to the direction of the profiles (b). The increase of the resistivity values on the deep part of the section clearly shows the interface between the volcanoclastic deposits and the underlying highly fractured carbonate bedrock. Figure 2c shows the bottom view of the volume data. In figure 2d, the pyroclastic horizons B, C, Bb and Bb$_{basal}$ are roughly marked.

## 3. Slope stability analysis through a geophysical Factor of Safety

The degree of slope hazard is usually expressed by the Factor of Safety (FS), which is defined as the ratio of resisting forces to disturbing forces (Fredlund and Rahardjo, 1993). Therefore, it follows that if $FS \leq 1$, disturbing forces are larger than resisting forces and failure may occur. Although the calculation of FS is a not easy task, since soils are physical systems where many components interact each other and with external perturbations, simplified analytical expressions of the Factor of Safety coming from the *infinite slope model* are often used (Haefeli, 1948; Taylor, 1948). Recently, several theoretical models based on an infinite-slope analysis have been developed to predict shallow landslide events (Sidle, 1992; Montgomery and Dietrich, 1994; Dietrich et al., 1995; Wu and Sidle, 1995; Baum et al., 2002; Savage et al., 2004; Baum et al., 2008). Such models provide an estimation of the FS values once geotechnical analyses have been performed on soil samples to determine physical and mechanical properties, such as friction angle, cohesion, soil and pore water unit weight and pressure head distribution.

* Corresponding author.
  *E-mail address*: esterpiegari@gmail.com (E. Piegari).

Alternatively to these approaches that need laboratory analyses on samples of reduced volume sizes, we attempt to describe the stability of a slope by analysing the distribution of in-situ measurable physical quantities, with the advantage to take into account the local changes of soil properties of considerable volumes. In previous papers (Piegari et al., 2006a; 2006b), we analyzed the probability of occurrence of a landslide event by using the Factor of Safety as the dynamical variable defining the state of the cells of a cellular automaton model. In particular, for the stability analysis of pyroclastic covers, in Piegari et al. (2009a) we introduced a semi-empirical expression for the local value of FS that depends on the slope angle, $\theta$, and on the electrical resistivity, through the following equation:

$$FS_i = \frac{\alpha}{\sin \theta_i}(\rho_i + \beta) \qquad (1)$$

where $i$ is the index that defines the $i-th$ cell of the grid, $\rho_i$ is the value of resistivity assigned to the cell $i$ coming from resistivity data inversion at the investigated depth below ground level, $\sin \theta_i$ is the maximum among the four slope angles that is possible to define for each cell, i.e. $\sin \theta_i = \max_j \{\sin \theta_{ij}\}$ with $\sin \theta_{ij} = (z_j - z_i)/\sqrt{(x_j - x_i)^2 + (y_j - y_i)^2 + (z_j - z_i)^2}$, where $(x_j, y_j, z_j, j=1,4)$ is the location of the four first neighbor cells of the cell $i$ ($j$=up, down, left and right), and $\alpha$ and $\beta$ are parameters which we determine imposing the stability thresholds. In detail, the two equations used to fix the values of $\alpha$ and $\beta$ read:

$$FS_{min} = \frac{\alpha}{\sin \theta_{max}}(\rho_{min} + \beta), \qquad (2)$$

$$FS_{max} = \frac{\alpha}{\sin \theta_{min}}(\rho_{max} + \beta). \qquad (3)$$

The first equation is the threshold condition for slope instability, where $\sin \theta_{max}$ is the largest slope angle measured in the study-area, (i.e. $\sin \theta_{max} = \max_i \{\sin \theta_i\}$), and $\rho_{min}$ is the smallest resistivity value coming from laboratory measurements on samples belonging to the same pyroclastic horizon in the saturated condition (De Vita et al., 2011; Di Maio and Piegari, 2011). As soon as the value of the Factor of Safety for a specific site $i$ (or a group of sites) reaches (or overcomes) the instability threshold $FS_{min}$, due to the action of external perturbations, the unstable site $i$ (or sites) relaxes and transfers its instability to neighbour cells. As a consequence of the relaxation process, the value of the Factor of Safety of the relaxed cells is fixed to the threshold $FS_{max}$, which is defined in terms of the smallest slope value, $\sin\theta_{min} = \min_i |\sin\theta_i|$, measured in the study-area and the largest resistivity value, $\rho_{max}$, coming from laboratory measurements on samples belonging to the same pyroclastic horizon in the dry condition (De Vita et al., 2011; Di Maio and Piegari, 2011). This is obviously not the only choice for $FS_{min}$ and $FS_{max}$. However, at the first order of approximation, our proposal represents the most cautious choice.

* Corresponding author.
 *E-mail address*: esterpiegari@gmail.com (E. Piegari).

It is worth noticing that Eq.(1) is a simplified expression that does not expect to work in general for slope stability analyses: indeed, it is well apparent that it does not work for instabilities correlated with high resistivity values (Piegari et al., 2009a). Since we are interested in landslide events triggered by rainfall infiltration and/or water outlets from bedrocks, i.e. by an increase of the water content in the field, at the first step of approximation the local value of the proposed Factor of Safety has a linear dependence on electrical resistivity.

### 3.1 Application to the test area on Mt. Pizzo d'Alvano

Let us now analyse the stability conditions of our survey area on the basis of Eq. (1). The nine transversal profiles of figure 1 delineate a grid area of $n = 225$ cells of size $(2 \times 4)\,\text{m}^2$, as resistivity measurements were performed at distances of 2 m along the cable and the nine parallel profiles are distant each other 4 m. We notice that the choice for the cell size is compatible with a previous analysis of the model, where an estimation of the maximum value for the cell size was obtained from the comparison of our landslide area synthetic probability distribution with that of three landslide inventories (Piegari et al., 2009b).

To attribute a value of FS to each cell of our grid area, the values of $\alpha$ and $\beta$ are needed. From Eqs. (2) and (3), it follows that

$$\alpha = \frac{FS_{max}\sin\theta_{min} - FS_{min}\sin\theta_{max}}{(\rho_{max} - \rho_{min})}$$

$$\beta = \frac{FS_{min}\sin\theta_{max}\rho_{max} - FS_{max}\sin\theta_{min}\rho_{min}}{FS_{max}\sin\theta_{min} - FS_{min}\sin\theta_{max}}.$$

The maximum and minimum slope angles of the grid cells are fixed from the topography of the area: $\sin\theta_{max} = 0.63$ and $\sin\theta_{min} = 0.0025$. The maximum and the minimum values of the electrical resistivity correspond to the dry and saturated conditions, respectively, and are obtained by laboratory measurements on samples collected in the same area (De Vita et al., 2011; Di Maio and Piegari, 2011). In detail, resistivity measurements were performed on fifteen samples belonging to three pyroclastic horizons: B horizon, at depths ranging from 0.4 m to 0.7 m, and Bb and Bb$_{basal}$, at depths ranging from 1.5 m to 2 m. The fifteen samples were saturated with rainwater (electrical conductivity σ=88μS/cm) at room temperature and standard pressure and, then, electrical measurements were performed decreasing the level of saturation by drying up the samples up to completely dry conditions (details of the experimental procedure are given in De Vita et al., 2011 and Di Maio and Piegari, 2011). For the shallowest pyroclastic horizon, it is found: $\rho_{max} \approx 5000\Omega m$ and $\rho_{min} \approx 250\Omega m$.

---
\* Corresponding author.
  E-mail address: esterpiegari@gmail.com (E. Piegari).

Having in mind that FS is the ratio of forces resisting movement to forces driving movement, which we try to express in terms of in-situ measurable (geo)physical quantities by means of Eq. (1), the instability threshold will set the value of $FS_{min}$ at 1. It follows that in the expression for α and β the only parameter to be fixed is $FS_{max}$. Although, in principle, no upper bound exists for such a quantity, we checked that our results do not significantly change in a wide range of values of $FS_{max}$, spreading on more than two orders of magnitude. However, to make a quantitative comparison with the Factor of Safety from the infinite slope model (see next section), we get the values of $FS_{max}$ from the following considerations.

Since we are interested in instabilities induced by rainfalls, which are linked to decreasing values of the resistivities, we consider only positive values for the constants α and β. By imposing $\alpha > 0$ and $\beta > 0$, it follows that $FS_{max} > \dfrac{\sin\theta_{max}}{\sin\theta_{min}} FS_{min}$ and $FS_{max} < \dfrac{\sin\theta_{max}}{\sin\theta_{min}} \dfrac{\rho_{max}}{\rho_{min}} FS_{min}$.

Such conditions limit $FS_{max}$ from about 250 to 5000. Moreover, from Eq.(1) it follows that $\alpha^{-1}$ has the dimension of resistivity. Therefore, we consider it as a normalization value, which we set equal to $\rho_{dry}$ measured on the pyroclastic samples collected in the same test area (Di Maio and Piegari, 2011). Such a constraint for the parameter $\alpha$ is observed if $FS_{max}$ ranges from 650 for the shallowest ashy layer to 500 for the deepest one.

Referring to the 3D resistivity maps of the investigated pyroclastic cover observed in the two seasons, we obtain the distributions of the geophysical Factor of Safety at different depths of investigation by using Eq. (1). Figure 3a shows the 3D resistivity distribution characterizing the cover in the autumnal season. In figures 3b, 3c, and 3d we report the corresponding distribution of the geophysical Factor of Safety at the depth of 0.70 m, 2.5 m and 4 m, respectively.

* Corresponding author.
  *E-mail address*: esterpiegari@gmail.com (E. Piegari).

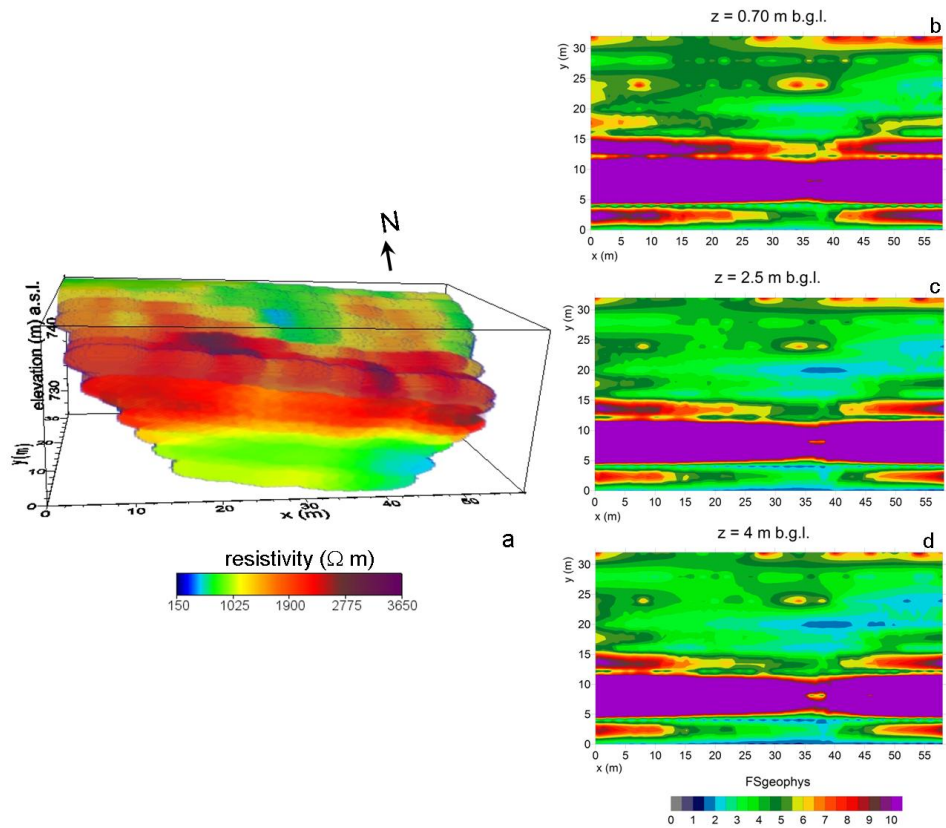

**Figure 3** a) 3D plot of the inverted ERT data obtained from the survey carried out in the test area (see figure 1) at the beginning of the autumnal season. b), c) and d) Maps of the geophysical Factor of Safety from Eq.(1), where the resistivity values characterize a thickness of the pyroclastic cover of about 0.70 m b.g.l. (b), the soils at about 2.5 m b.g.l. (c) and at about 4 m b.g.l. (d).

As it can be seen, in the southern part of the site along the whole horizontal axis, a very stable area appears (located approximately at coordinates 5-10 m on the vertical axis): this is due to the presence of a terrace, which implies very small slope angle and, consequently, very large values of the Factor of Safety at all depths of investigation (see figures 3b, c and d). Actually, the values of FS ranges from about 1.1 to 200 and, therefore, in the color scale of figures 3 we saturate the values larger than 10 for better visualization of data. However, a weak point in such a stable area is visible at distances ranging approximately between 35-40 m on the horizontal axis, which it is better detectable as the depth is increased. Interestingly, such a point of weakness is in correspondence of a step in the underlying bedrock structure, which is indicated with an ellipse in figure 2c. We notice that, the point of weakness is offset from the bedrock step in figure 2 just for an effect of the 3D view of the investigated volume, which prevents the correct estimation of the step location. In addition, the lowest values of the Factor of Safety are found exactly above and below such a point of weakness, at all investigated depths, i.e. z coordinates (see sections 3 and 4), revealing the most unstable conditions in correspondence of a buried structural discontinuity.

* Corresponding author.
 *E-mail address*: esterpiegari@gmail.com (E. Piegari).

To study the effect of rainy winter days on the stability of the cover, in figure 4a we report the 3D resistivity distribution measured at the beginning of the spring season, and in figures 4b, 4c and 4d, we report the corresponding distribution of the geophysical Factor of Safety at the depth of 0.70 m, 2.5 m and 4 m, respectively. The values of FS ranges from about 1.03 to 150 and in the color scale of figures 4 we saturate the values larger than 10 for better visualization of data. As expected, the winter season has the effect of enlarging the region characterized by the smallest values of FS and shrinking the very stable area. It is worthwhile noticing that even if significant variations of resistivity values are found in the uppermost soil layer (see figures 3a and 4a), such variations essentially affect the very stable area and, therefore, they do not produce a significant decrease of the FS values, which are obtained by the ratio between resistivity values and very low slope angles. Anyway, the biggest differences between the FS maps in the autumnal and spring season are found for the upper layer (figures 3b and 4b, respectively), while the geophysical FS maps are more similar as the depth is increased (figures 3c,d and figures 4c,d). We interpret such a result as a consequence of the large capacity of these materials to retain water due to their large porosity values (Di Maio and Piegari, 2011).

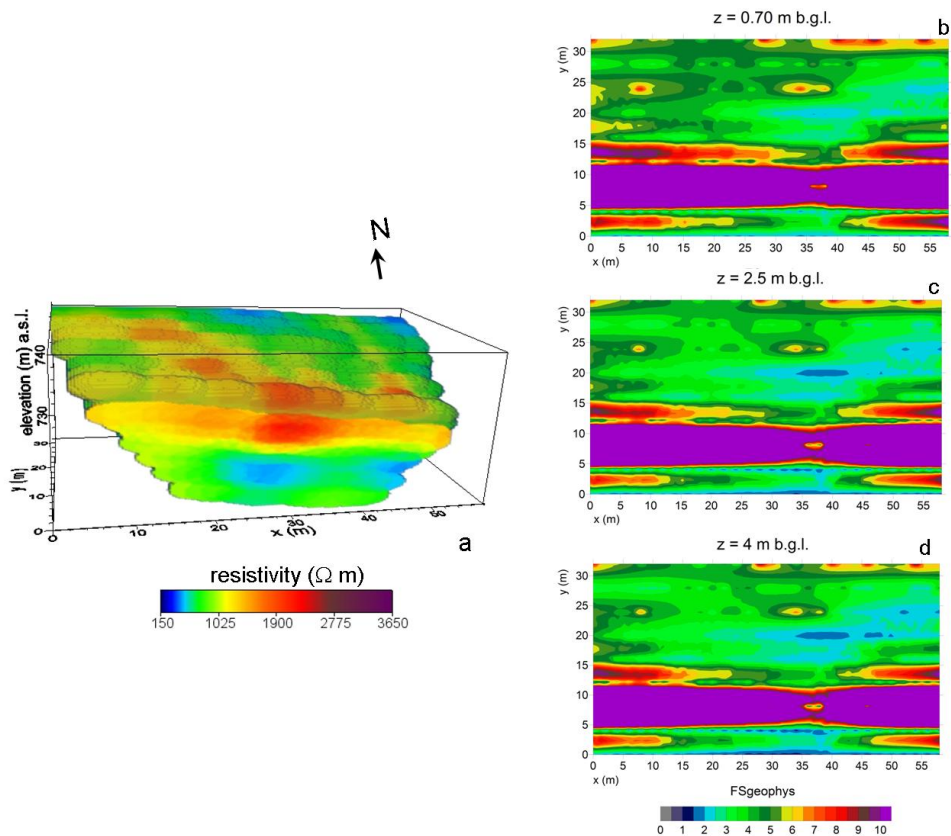

**Figure 4** a) 3D plot of the inverted ERT data obtained from the survey carried out in the test area (see figure 1) at the beginning of the spring season. b), c) and d) Maps of the geophysical Factor of Safety from Eq.(1), where the resistivity values characterize a thickness of the pyroclastic cover of about 0.70 m b.g.l. (b), the soils at about

2.5 m b.g.l. (c) and at about 4 m b.g.l. (d).

## 4. A comparison of the geophysical stability analysis with the infinite slope model

Nowadays, the simplest model used for analyzing slope stability is the infinite slope model, which assumes the failure plane and water table parallel to the ground surface. Although such a model relies on some very simplifying assumptions, it is often used in conjunction with hydrological models to forecast shallow landslides induced by rainfalls (Montgomery and Dietrich, 1994; Iverson, 2000; Baum et al., 2002; Savage et al., 2004). In particular, for the back-analyses of May 1998 shallow landslides along the Pizzo d'Alvano slopes, three different types of models have been used with an infinite slope stability analysis for the computation of the Factor of Safety (Sorbino et al., 2010).

In the following, we compare the values of the Factor of Safety from our empirical model, Eq.(1), with those coming from the infinite slope model. In this latter case, the formula to calculate the Factor of Safety (Brunsden and Prior, 1984) reads:

$$FS = \frac{c + (\gamma - m\gamma_w)z\cos^2\vartheta\tan\phi}{\gamma z \sin\vartheta\cos\vartheta} \quad (4)$$

where $c$ is the effective cohesion (Pa), $\gamma$ is the unit weight of soil (N/m$^3$), $\gamma_m$ is the unit weight of water, $m = z_w/z$ is the ratio between the height of watertable above failure surface ($z_w$) and the depth of failure surface below the surface ($z$), $\phi$ is the effective angle of shearing resistance (°), $\vartheta$ is the slope surface inclination (°). Eq. (4) can be easily applied for various heights of the watertable ($z_w$). We study two limit cases: the dry scenario, which is defined by the condition $z_w=0$ and, therefore, $m=0$, and the saturated scenario, when all the soil above the failure surface is water saturated, i.e. $z_w=z$, and, therefore, $m=1$. To calculate the values of the geotechnical Factor of Safety given by Eq.(4), we use the physical and mechanical parameters provided by accurate laboratory analyses on ashy soil layers of the Sarno Mountains performed by Cascini et al. (2003): $\gamma_{dry}=7.3kN/m^3$; $\gamma_{sat}=13.1\ kN/m^3$; $c=4.7\ kPa$; $\phi=32°$ for upper ashy soil layer, and $\gamma_{dry}=9.10kN/m^3$; $\gamma_{sat}=15.7\ kN/m^3$; $c=4.7\ kPa$; $\phi=32°$ for lower ashy soil layer. Substituting such values in Eq.(4) and using the values of the slope angles as above (see section 3), we obtain the maps of the geotechnical Factor of Safety for the dry and saturated scenario, by considering the values of the failure surface depth: $z = 0.70$ m for upper ashy layer (figures 5a and 5b), and $z = 2.5$ m and $z = 4$ m for lower ashy layer (figures 5c, 5d, 5e and 5f). As expected, a large stable area appears in the dry scenario for $z = 0.70$ m (figure 5a), which narrows significantly in the saturated scenario (figure 5b), including the only southern part of the site. In particular, for such a scenario, if the failure surface is at depths of $z = 2.5$ m and $z = 4$ m (figures 5d and 5f), values of FS less than 1 are found above and below the stable terrace.

Let us now compare the results of FS obtained with the two different methods. First, we compare the maps of the geophysical Factor of Safety at the beginning of the autumnal season (figures 3b, 3c and 3d) with those of the dry scenario (figures 5a, 5c and 5e). We notice that the distribution of the geotechnical Factor of Safety in the dry scenario with the failure surface at $z = 0.70$ m (figure 5a) is quite different from that of the geophysical Factor of Safety at the same depth of investigation (figure 3b). Such differences are expected if one considers the effect of small water content in the field immediately after the summer season, which the infinite slope model in the dry scenario does not take into account. Instead, the maps of the geophysical and geotechnical FS appear more similar if larger depths are considered. On the other hand, if we compare the results for the geophysical FS in the spring season (figures 4b, 4c and 4d) with those of the geotechnical FS in the saturated scenario (figures 5b, 5d and 5f), the largest differences are found as the failure surface depth increases. In particular, the saturated scenario with a failure surface at $z = 2.5$ m predicts large sectors with FS values close to the instability threshold, and at $z = 4$ m forecasts sectors with FS below the instability threshold. However, instabilities with FS < 1 are found in the same areas where the geophysical FS has the lowest values, even if larger than 1.

Summarizing, the proposed geophysical approach to stability analysis seems to provide results comparable with those from the infinite slope model. In particular, geophysical FS maps correctly locate the areas that are more susceptible to sliding and give information on the distance from the limit cases of the infinite slope model. Specifically, geophysical analyses reveal the presence of a potential unstable area at a depth of 0.70 m b.g.l., which is the typical soil thickness for landslide triggering on Sarno Mountains (Cascini et al., 2003), with map coordinates 35-40 m on the *x* axis and 15-20 m on the *y* axis, already at the end of the summer, as shown in figure 3b. The size of such a potential unstable area enlarges during the spring season, as reasonable expected.

* Corresponding author.
 *E-mail address*: esterpiegari@gmail.com (E. Piegari).

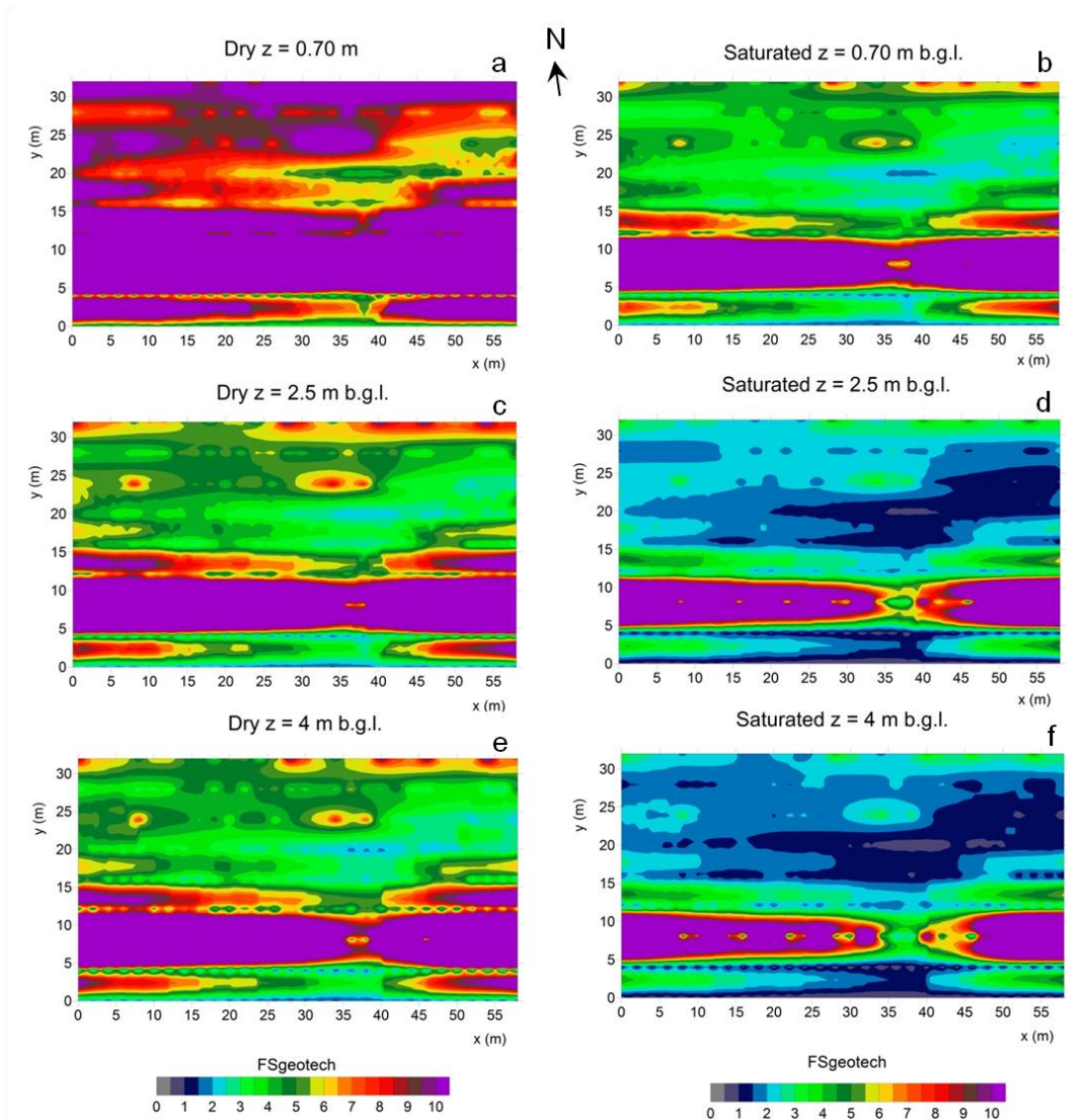

**Figure 5** Maps of the Factor of Safety from the infinite slope model. Dry (a) and saturated (b) scenario considering a failure surface depth of 0.7 m. Dry (c) and saturated (d) scenario considering a failure surface depth of 2.5 m. Dry (e) and saturated (f) scenario considering a failure surface depth of 4 m.

## 5. Discussion and conclusions

Rainfall is the most frequent trigger of landslides involving ash-fall deposits covering the mountains close to the Mt. Somma-Vesuvius volcanic complex. It follows that the water content of such soils is a key element for triggering instability and its knowledge is fundamental to determine reliable warning thresholds. To overcome the limit of local information provided by rain gauges, we propose the use of the electrical resistivity tomography (ERT) technique for stability analyses of selected areas. It is worth to point out that the use of ERT to investigate landslides is widely applied (Jongmans and Garambois, 2007). Recently, in fact, automated resistivity tomography systems with permanently installed electrode networks have been developed to monitor the hydraulic precursors to landslide movement (Supper et al., 2008; Chambers et al., 2009; Lebourg et al., 2009; Wilkinson et al., 2010). In particular, in this paper we propose to relate time-lapse resistivity changes to slope failure by introducing a geophysical Factor of Safety, which depends on resistivity data and site topography (Piegari et al., 2009a). The idea to assume an explicit dependence of the FS on in-situ measurable geophysical quantities has two main advantages. First, it enables susceptibility variations to be related to soil property changes in large volumes, overcoming the limit of point information provided by rain gauges and porous probes. Second, it allows to monitor the slope stability through noninvasive techniques, which do not alter the soil structure, contrary to classical measurements that perturb the soil by drilling and sampling. Even if our approach needs further validations, limitations appear in the simple expression of the Factor of Safety which exclusively links instability to decreasing resistivity values. For this reason, we only consider the case of susceptibility to sliding induced by rainfalls and apply our model to stability of pyroclastic soils, which have been previously characterized by many accurate geotechnical analyses (Calcaterra et al., 2003; Cascini et al. 2003, De Vita et al., 2006). Specifically, we used the 3D inversion results of two 2D ERT surveys carried out, respectively, at the beginning of both the autumnal season (at the end of September) and the spring season (at the end of March) to compute the values of the proposed geophysical Factor of Safety. The maps of the geophysical FS were compared with those derived from the infinite slope model in the dry and saturated scenario. The similarities between the maps of the geophysical and geotechnical Factors of Safety show the potentiality of the proposed approach, being the validity of the infinite slope model recognized by previous studies (e.g. Sorbino et al., 2010).

* Corresponding author.
  *E-mail address*: esterpiegari@gmail.com (E. Piegari).

We point out that our research is ongoing and we aim to expand the expression of the geophysical Factor of Safety by taking into account the dependence on different geophysical quantities (e.g. seismic wave velocity), which are also related to soil saturation. However, at this stage of approximation, we are searching for linking resistivity changes to rainfall, in order to estimate reliable empirical thresholds. At present, in fact, we are looking for a possible procedure to calculate the rain amount and duration needed to trigger a slope failure in our test area.

## ACKNOWLEDGMENTS

We wish to acknowledge P. De Vita, S. Di Nocera and F. Matano for useful discussions.

* Corresponding author.
  E-mail address: esterpiegari@gmail.com (E. Piegari).

\* Corresponding author.
   *E-mail address*: esterpiegari@gmail.com (E. Piegari).

* Corresponding author.
 *E-mail address*: esterpiegari@gmail.com (E. Piegari).

\* Corresponding author.
 *E-mail address*: esterpiegari@gmail.com (E. Piegari).